\documentstyle[12pt]{article}
\begin{document}
\vspace{4.5cm}
\centerline {\large \bf UNIFIED MICROSCOPIC THEORY OF A SYSTEM OF}
\centerline {\large \bf INTERACTING BOSONS}
\vspace{1.7cm}
\centerline {{\bf Y.S. Jain}\footnote{E-mail:jain@nehus.ren.nic.in}}
\vspace{.3cm}
\centerline {Department of Physics, } 
\smallskip
\centerline {North-Eastern Hill University, Shillong-793 022, India} 
\vspace{4cm}
\begin{abstract}
This paper reports the unified microscopic theory
 of a system of interacting bosons, 
such as liquid $^4He$.  Each particle in the system 
represents a $({\bf q}, -{\bf q})$ pair moving with a centre of
 mas momentum {\bf K}.  Particles form bound pairs below $\lambda$-
 point and have a kind of collective binding between them.  The 
 binding is identified as an energy gap between 
 the superfluid and normal states of the system.  
  The $\lambda$-transition is a consequence of inter-particle
 quantum correlations.  It follows an order-disorder of 
particles in their phase space as well as the onset of Bose Einstein
 condensation in the state of $q={\pi}/d$ and $K=0$.  In addition 
 to the well known modes of collective motions such as phonons, rotons, 
 maxons, {\it etc.}, the superfluid 
 state also exhibits a new kind of quantum quasi-particle, {\it omon}, 
 characterised by a phononlike wave of the oscillations of momentum 
 coordinates of the particles.  The theory explains the properties 
 of $He-II$ at quantitative level and vindicates two fluid theory
 of Landau.  The paper, finally, describes the way this theory 
could help in understanding the superfluidity of 1-D and 2-D  
 systems.  It also analyses the possibility of applying 
this approach to develop similar framework for  
a fermion system including an atomic nucleus.               
\end{abstract}
\vspace{20cm}
\centerline{\bf 1. INTRODUCTION}
\bigskip
A system of
interacting bosons, such as Liquid $^4He$, forms an
 important subject of
study for its low temperature $(T)$ behaviour.  Liquid $^4He$
transforms from its normal (N) phase $(He-I)$ to superfluid (S) phase
$(He-II)$ at $T_{\lambda} = 2.17 K$; the latter shows 
zero viscosity ($\eta = 0$) when it flows 
through narrow channels.  
The phenomenon has been investigated extensively because   
it provides a unique opportunity to study quantum
behaviour of a system at macroscopic level.  Different aspects of the
subject  have been reviewed in several articles
and books, {\it e.g.} [1-8].  Widely varying ideas and 
mathematical tools have been
used to develop possible microscopic theories of the system but   
the desired theory, that explains 
the properties of liquid ${^4He}$, could 
not emerge.  As per the contemporary belief superfluidity arises due to
the existence of zero momentum $(p=0)$ condensate, $n(o)$, representing 
a macroscopically large fraction of particles having Bose
Einstein  Condensation (BEC) in a single particle state of $p=0$.
However, this premise could also not help in 
achieving the goal.  Analysis of certain experimental results reveals 
$n(o) \approx 0.12$ [6] but not with unquestionable certainty .  This 
motivated us to develop the present theory 
by using an entirely new approach to the
problem.  Here we present the
salient aspects of our theory [9,10].
\vspace{.6cm}
\centerline {\bf 2.BASIC ASPECTS OF THEORY}
\vspace{.3cm}
\centerline {\bf 2.1 The System}
\vspace{.3cm}
The Hamiltonian of the system can be expressed as 
$$ H(N) = - \frac{{\hbar}^2}{2m} {\sum_i^N} {\bigtriangledown}^2_i + 
{\sum_{i<j}^N}
V_{ij}(r = |{\bf r}_i - {\bf r}_j|). \eqno(1)$$
\noindent
where notations have their usual meaning.  To a good approximation
 $V_{ij}(r)$ is the sum of : (i) the hard core (HC) repulsion
(i.e. $V_{ij}(r<\sigma) = \infty$ and $V_{ij}(r\ge \sigma) = 0$
with $\sigma$ = HC diameter), and (ii) a relatively long range weak
attraction.  Since the latter can be replaced by a constant negative
external potential, we have only the HC to deal with.  We note that two
particles of momenta ${\bf k}_1$ and ${\bf k}_2$ (as seen in
the laboratory frame) see each other as particles of equal and opposite
momenta of the value, 
$k$ = $|{\bf k}_1 - {\bf k}_2|$.  In fact in 
a frame attached to their centre of mass (CM),
their momenta are found to be {\bf q} 
and -{\bf q} with {\bf q} = {\bf k}/2.  Naturally, 
a state function of the system should be in
conformity with this basic fact.  To this effect we first examine 
the wave mechanics of two HC particles.  
\vspace{.3cm}
\centerline {\bf 2.2 Wave Mechanics of Two HC Particles}
\vspace{.3cm}
In the CM coordinate system a pair of HC particles is described by 
$$ [({\hbar}^2/{4m}){\bigtriangledown}^2_R + ({\hbar}^2/{m}) 
{\bigtriangledown}^2_r + V_{HC}(r)]{\Psi}(R,r) =
E{\Psi}(R,r). \eqno(2)$$
\noindent
The notations $R$, and $r$ have their usual meaning through ${\bf R} =
({\bf r}_i + {\bf r}_j)/2$, and ${\bf r} = ({\bf r}_i - {\bf r}_j)$. 
${\Psi}(R,r)$ has a form  ${\psi}_k(r)\exp(i{\bf K}.{\bf R})$
with {\bf K} being the CM momentum.  The form 
of ${\psi}_k(r)$ describing the relative
motion of two particles, depends on the way we deal
 with $V_{HC}(r)$ in solving
Eqn.(2).  While other studies [11] use a
 boundary condition, {\it i.e.} ${\psi}_k
(r<{\sigma}) = 0$ and ${\psi}_k(r \geq {\sigma}) \ne 0$ or its equivalent
({\it e.g.} Jastrow type correlation [12]), we use a new condition.
Accordingly, the
separation ($d$) between two HC particles should
satisfy $d {\ge} {\lambda}/2 \quad({\rm with} \quad \lambda$ = $2\pi/q$).
 In this context, we note that a
particle  in wave mechanics manifests 
itself as a wave packet (WP) of size
${\lambda}/2$; as two HC particles do not overlap, their WPs
should do likewise.  If two particles, somehow, have 
${\lambda}/2 > d$ they experience mutual 
repulsion until increased $d$ satisfies 
 $d = {\lambda}/2$; in case the system does not allow desired
increase in $d$, particles would absorb necessary energy from the
interacting surroundings so that $\lambda/2$ gets
 squeezed to satisfy this
condition.  Since two particles, in a physically 
possible state,
always have $r \ge \sigma$, {\it i.e.}  $V_{HC}(r)$ = 0, 
the solutions, ${\Psi}(R,r)$, of 
Eqn.(2) take the form of $U^{\pm}(R,r) \equiv U^{\pm}$.  We have 
$$ U^{\pm} = ({\sqrt 2}).\cos[(\alpha + {\bf
k.r})/2].\exp[i{\bf K}.{\bf R}].\exp[-i{(\varepsilon{(K)} + 
\varepsilon{(k)})}t/{\hbar}]  \eqno(3)$$
\noindent
with $E= \varepsilon{(K)}+ \varepsilon{(k)}=
  \hbar^2[K^2+k^2]/4m = \hbar^2[k_1^2 +k_2^2]/2m$, $\alpha = 0$ 
for $U^+$ and $\alpha = 
\pi$ for $U^-$.  A critical 
 analysis of $U^{\pm}$ reveals that : 
\smallskip
\noindent
{\bf (a).} The probability that two particles have a phase
separation $\phi = \bf k.r$ is 
$$|U^{\pm}|^2 = [1 + \cos{(\alpha + \bf k.r)}]
 = [1 + \cos\phi']
= {|{\psi}_k(r)^{\pm}|}^2   \eqno(4)$$
\noindent
Since  $|U^{\pm}|^2$ is independent of ${\bf K}$, the relative
configuration of the pair can be derived by using ${\bf K} = 0,$
{\it i.e.} ${\bf k}_1 = -{\bf k}_2 = {\bf q}$,  without loss of
generality.  Eqn.(3) gives ${\psi}_k(r)^{\pm} \equiv  {\psi}^{\pm} = 
 {\sqrt {2}}.\cos[(\alpha + {\bf
k.r})/2].\exp[-i.{\varepsilon{(k)}}t/{\hbar}]$ that represents a   
(${\bf q}$, -${\bf q}$) pair.  As $ {\psi}^{\pm}$ 
is a kind of stationary matter wave (SMW)
that modulates the probability of finding two particles in $\phi$
space, the $(\bf q, -\bf q)$ pair can be known as SMW pair. 
 Note that $ {\psi}^+$  and $ {\psi}^-$  
differ only in the locations of the origin
${\phi}' = 0$.  For $ {\psi}^+$, it is located at  
 $\phi = 0$, -the central point of
an antinodal region (AR), while for $ {\psi}^-$ at $\phi = \pi$ 
(the nodal point) of a SMW.  Obviously, $ {\psi}^+$  and $ {\psi}^-$
are equivalent.      
\smallskip
\noindent   
{\bf (b).} Each AR of $ {\psi}^{\pm}$ is 
of ${\lambda/2}$ size and $<r>$ of a particle
evaluted over an AR is found to be its central
point.   This  gives $<r_1> - <r_2> = \lambda/2$ if we assume
 that each AR is occupied, exclusively, by one particle.  The
 fact, that this configuration ensures no overlap of two WPs,  
 implies that particles satisfying ${\lambda}/2 \le d$ render 
 $<V_{HC}(r)> = 0$.  
\smallskip
\noindent
{\bf (c).}  Since each particle of the pair is a 
representative of $ U^{\pm}$ state it needs to be expressed by 
$U^{\pm}$.   In this context we note that
a particle in a closed enclosure has two fields: $u_{k'}(r') =
(1/{\sqrt {\rm V}}).\exp(i{\bf k'.r'})$ and 
 $u_{k''}(r'') = (1/{\sqrt {\rm V}}).$  $\exp(i{\bf k''.r''})$;
 the latter is
the reflection of the former.  To account for the reflected fields, we
arrange 
$$H(2) =  -{1\over 2}\sum^2_{j=1}{{\hbar}^2\over
2m}\left[{\bigtriangledown}^2_{r'_j}  +
{\bigtriangledown}^2_{r''_j}\right] = - {1\over
2}\sum^2_{j=1}\left[{{\hbar}^2\over  4m}
{\bigtriangledown}^2_{R_j} +
{{\hbar}^2\over m}{\bigtriangledown}^2_{r_j}\right]. \eqno(5)$$
\noindent
A $U^{\pm}$ of j-th particle can be the superposition of any two
of the four $u_k(r)$.  The choices are :$(1'1''), (1'2''), (1'2'),
(2'1''), (2'2'')$,  and $(1''2'')$  with numbers
referring to particles in superposition.  The possible state functions
rendered by these $U^{\pm}$ are: $\Psi_2^{(1)} = (1'1'')(2'2'')$,
$\Psi_2^{(2)} =  (1'2'')(2'1'')$ and  $\Psi_2^{(3)} =  (1'2')(1''2'')$. 
\smallskip 
\noindent
{\bf (d).}  A pair waveform such as $(1',1'')$ is a case of
 self superposition
(SS) of a particle where $V_{HC}(r)$ does not operate; naturally, $U^+$ can be
used rightly to represent the situation.  But the other cases such as
$(1',2')$, $(1',2'')$, etc.  represent mutual superposition (MS) of two
particles.  The situation is best represented by $U^-$ because the
waveform is expected to vanish at $r = 0$.  However, in a state of wave
mechanical superposition (i.e. $ \lambda \ge 2\sigma)$, there is no
way to determine whether two particles have SS or MS.  But then
 we use either $U^+$
or $U^-$ for all particles to keep a single frame of observation. 
Use of $U^-$ for two HC bosons is consistent with the fact that such
bosons do not have common $r$ coordinates and they behave like fermions in
this respect.  In momentum space, however, while any number of bosons can have
equal $K$, two identical fermions have different values of $K$.    
\bigskip
\centerline{\bf 2.3 State Functions of $N$ HC Particles} 
\bigskip
Following the arguments of Section (2.2) for taking care 
of $V_{HC}(r)$, we can express $H(N)$ the way Eqn.(5) expresses $H(2)$, 
and describe each
particle by a pair waveform $U^{\pm}$.  
 For $N$ particles, we 
 have $2N$ plane waves and $N(2N-1)$ different $U^{\pm}$
rendering $S$ = $1.3.5...(2N-1)$ different $\Psi_N$ (state functions
of equal $E$).  Using $E(K) = \sum_i^N{\varepsilon{(K)_i}}$
 and $E(k) = \sum_i^N{\varepsilon{(k)_i}}$,  we have 
$$\Psi_N =  \phi_N(q).\phi_N(K).  \eqno(6)$$ 
$$\phi_N(q) = \left[ \left( \frac{2}{\rm V} \right) ^{\frac{N}{2}}
\prod_{j=1}^{N} \cos (\alpha + 2{\bf q}_j.{\bf r}_j)/2
\right]\exp[-iE(k)t/{\hbar}].  \eqno(6a)$$    
$$\phi_N(K) = \left[{A}.{\left( \frac{1}{\rm V} \right)}^{\frac{N}{2}}
{\sum_{{\it p}K}}{(\pm 1)^p} \prod_{j=1}^{N}
 \exp [i({\bf K}_j.{\bf R}_j] \right]\exp[-iE(K)t/{\hbar}]
 \eqno(6b)$$
\noindent
with $A={\sqrt {1/N!}}$.  Here $\sum_{{\it p}K}{(\pm 1)^p}$ refers
 to the sum of different permutations of $K$ over all
particles.  While the choice of $(+1)^p$ or $(-1)^p$
 in Eqn.(6b) distinguishes between the systems of bosons
 and fermions, use of the restriction $q_j \ge \pi/d$ in Eqn.(6a) 
 treats the fermion type behaviour due to HC nature of bosons
 and fermions alike .  $S$ different 
$\Psi_N$ counted above take care of the
 permutation of $k$.  We have 
$$\Phi_N =  \frac{1}{\sqrt S}.\sum_i^{S} \Psi_N^{(i)}. \eqno(7)$$
\noindent
which represents the general form of a state
 function that should reveal 
the physics of the system.  Note
 that ${\Phi_N}$ represents a state where each particle 
(as a WP of size $=\pi/q$) has a plane wave motion 
of momentum {\bf K}  and in conformity with our expectation ({\it cf.}
Section 2.1) truely has
$({\bf q}, -{\bf q})$ configuration. As such a particle has two
 motions: (i) the $q$ motion of energy $\varepsilon{(q)}$
 = ${\hbar^2q^2}/2m$, and (ii) the 
$K$ motion of energy $\varepsilon{(K)}$= ${\hbar^2K^2}/2(4m)$.
\bigskip
\centerline{\bf 2.4 Ground State Energy} 
\bigskip
The condition $\lambda/2 \le d$ implies that a spherical volume of diameter 
$\lambda/2$ belongs exclusively to an HC particle of 
$\lambda/2 > \sigma$.  We also note that each particle in the ground state has
lowest possible energy, ${\it i.e.}$ largest possible $\lambda/2$, the net
ground state energy of the system should be 
$$E_o = \sum_i^N \frac{h^2}{8m{{\rm v}_i^{2/3}}} \quad \quad {\rm and}
 \quad \sum_i^N
{\rm v}_i = {\rm V} \quad {\rm (constant)}   \eqno(8)$$
if particles are assumed to occupy different ${\rm v}_i$.  
  Simple algebra reveals that $E_o$ has its minimum value for $ {\rm v}_1
 = {\rm v}_2 =.. {\rm v}_N = {\rm V}/N$. Obviously, 
$$E_o = Nh^2/8md^2 = N{\varepsilon}_o    \eqno(9)$$
In sharp contrast with 
$E_o$ obtained from convesional theories [11, 12] 
our $E_o$ does not depend on $\sigma$.  The 
accuracy of this aspect of our result is well evident because 
two particles of $\lambda/2 > \sigma$ cannot resolve the HC 
structure within the larger size WPs of each other.
 The belief in such possibility would 
contradict the basic principle of image resolution. 
\bigskip
\centerline{\bf 2.5 Evolution of the System with Decreasing $T$} 
\bigskip
For constant particle density $(d-{\lambda}/2)$ 
decreases with decreasing $T$.  In the process at
certain $T = T_c$, when  $d-\lambda/2$ vanishes at large, $q$ motions
get freezed into zero point motions of $q = q_o = \pi/d$.
Evidently, the system moves from a
state of $\lambda/2 \le d$ to that of $\lambda/2 = d$.  While the former
state of $\phi {(=2qd)} \ge 2\pi$ represents randomness of $\phi$
positions and, therefore, a disorder in $\phi$ space, the latter of
 $\phi = 2\pi$ defines an ordered state.  Thus the
particles in the system move from their disordered to ordered state in
$\phi$ space at $T_c$.  With all $q_j = q_o$, different  $\Psi_N^{(i)}$
 of Eqn.(7) become identical
and $\Phi_N$ attains the form of a single $\Psi_N$ (${\it cf.}$ Eqn.6). 
As such all the  $S$
microstates merge into one at $T_c$ indicating that the entire system 
attains a kind of oneness [13]; the system at $T\le T_c$ is, 
 therefore, described by
$$\Phi_N(S) = \phi_N^o(q_o).\phi_N^e(K) \eqno(10)$$ 
obtained by replacing all ${\bf q}_j.{\bf r}_j$ in
 Eqn.(6) by $q_or$ as for given $r$ and ${\bf q}.{\bf r}
 = \pi$ ({\it i.e.} $r\cos{\theta} =\lambda/2$), the lowest energy
 configuration demands $\theta =0$.  
\bigskip
\centerline{\bf 3. QUANTUM CORRELATION POTENTIAL} 
\bigskip
The interparticle quantum correlation potential (QCP), 
originating  from the wave nature of
particles, is obtained by comparing the
 partition function (under the quantum limits of the system), 
 $Z_{\rm q} =
 {\sum}_n \exp(-E_n/k_BT).|{\Phi}_n|^2$ and its classical
 equivalent, $Z_{\rm c} =
 {\sum}_n \exp(-E_n/k_BT).\exp(-U_n/k_BT)$.  Here ${\Phi}_n$ is
 a $\Phi_N(S)$ (Eqn.(10)) of $n$-th 
state.   The procedure is justified because our theory describes 
the system by symmetrised plane waves and our assumption that 
only one particle occupies a single AR of these SMWs    
screens out the HC potential.  Simplifying $U_n$, one
  easily finds that pairwise  
 QCP has two components [9].  The $U^s_{ij}$
  pertaining to $k$ motion 
 controls the $\phi = kr$ position of
 a particle and we have  
$$U^s_{ij} = - k_BT_o.\ln[2\cos^2({\phi}'/2)] \quad \quad \quad {\rm with}
\quad \quad {\phi} '= (\alpha +\phi),  \eqno(11a)$$ 
\noindent
where $T$ has been replaced by $T_o$ because $T$ equivalent of $k$
motion energy at all $T \le T_{\lambda}$ is $T_o$.      
\smallskip
  $U^s_{ij}$ has its minimum value 
 $(-k_BT_o.\ln{2})$ at ${\phi}' = 2n{\pi}$ and maximum value 
  $(=\infty)$ at ${\phi}' = (2n+1){\pi}$ occurring periodically
at $\Delta\phi = 2n{\pi}$ (with $n = 1,2,3,...$).  Since $U^s_{ij}$
increases as ${\frac{1}{2}}.C({\delta\phi})^2$ ( with $C =
 {\frac{1}{2}}k_BT_o)$ for 
small change $\delta\phi$ in $\phi$
around a point of its minimum, it generates a force = -$C{\delta\phi}$
which tries to maintain $\delta\phi = 0$ and hence the order of particles
in $\phi$ space.  Since $U^s_{ij}$ is
not a real interaction, in general, it cannot manipulate $d$.  The
$\phi$-space order is, therefore, achieved by driving all $q$ towards
$q_o$.  In the S-phase, however, the interdependence of $r$ and 
$q$ through $\phi = 2qr = 2n\pi$ renders $V_{ij}$ a function
 of $\phi$ and $q$ values.  More accurately ({\it cf.} Section 5.4),
 particles in S-state acquire self energy (a kind of additional potential 
energy) that depends on $q$ values.  Evidently, $V_{ij}$ in 
S state can be replaced by $U^s_{ij}$. 
\smallskip
 The second component pertaining to $K$ motion is expressed by   
$$U_{ij} = -k_BT.\ln{[1} + {\exp{(-2{\pi}{|{\bf R'}-{\bf R''}|^2})}/
{\lambda}'^2_T]}, \eqno (11b)$$
\noindent 
 with ${\lambda}'_T = h/{\sqrt{2\pi(4m)k_BT}}$.  $U_{ij}$ may
 be seen as the origin of a force that facilitates 
 BEC in the state of $K = 0$ by driving particles in $K$ space 
 towards $K =0$ where it has its minimum value ($- k_BT.\ln{2}$). 
\bigskip
\centerline{\bf 4. THE  TRANSITION} 
\bigskip
To a good approximation, the lower bound
 of $T_c$ is $T_o$ (the $T$ equivalent of 
$\varepsilon_o$ or that of $\lambda = 2d)$. This gives 
$T_c$ =  $T_o$ = $h^2/8{\pi}mk_Bd^2$, by using 
${\lambda}_T = h/{\sqrt{(2{\pi}mk_BT)}}$.  
 To reveal the real $T_c
 = T_{\lambda}$, we note that with $T$ moving below $T_{\lambda}$ 
 particles not only have $q$ = $q_o$, but also start
attaining $K = 0$.  Evidently, the ${\lambda}$-transition follows
two processes simultaneously: (i) an order-disorder of particles
 in $\phi$ space rendering $q=q_o$
 and (ii) the BEC
of particles (as SMW pairs) driving them towards      
 $K= 0.$  In view of these aspects it is evident that 
 $\lambda$-transition 
is a second order transition.
  This also gives  
 $$T_{\lambda} = T_o + \frac{1}{4}.T_{\rm BEC} =
 \frac{h^2}{8{\pi}mk_B}.{\left[{\frac{1}{ d^2}} +
{\left({\frac{\rm N}{2.61{\rm V}}}\right)}^{\frac{2}{3}}\right]}
  \eqno (12)$$  
\noindent
where $T_{\rm BEC}$ is usual BEC
 temperature [4].  The factor of $\frac{1}{4}$ appears because 
the plane wave $K$ motion of a particle has 
 ${\hbar}^2{K}^2/2(4m)$ energy and  
$T_{\rm BEC}$ varies as $\frac{1}{m}$.    
The sharpness of the transition is well evident.      
\bigskip
\centerline{\bf 5. PROPERTIES OF S-PHASE}
\bigskip
\centerline{\bf 5.1 Configuration of Particles} 
\bigskip
Our key results, $\Delta\phi =2n{\pi}$ and $q = q_o$, revealing
 uniformally equal $d$,
 define $\phi$, $q$, and $r$ space
 configurations.  The system can be identified as a close packed 
arrangement of WPs.
The packing leaves no freedom for particles to move across
each other.  They can move in the order of their locations 
maintaining $\Delta\phi$= $2n\pi$.  Naturally, the
motion has coherence.  
\bigskip
\centerline{\bf 5.2 Thermal Excitations.} 
\bigskip
Since $U^s_{ij}$ restores $\phi = 2n\pi$ and the system is 
a close packed arrangement of WPs, we 
visualise waves  
of $\phi$ oscillations. To reveal their frequency dispersion,
${\omega}_{\phi}{(\rm Q)}$, 
we consider a linear chain of atoms and
 only nearest neighbour interactions 
as the responsible forces.  We have  
$${\omega}_{\phi}{(\rm Q)} =
 {\sqrt {(4C)/{\beta}}}.|\sin({\rm Q}d/2)| \eqno (13)$$
\noindent
where Q is the wave vector and $\beta$ is the measure
 of inertia for ${\phi}$ motion.  However, 
 $\phi$ oscillations can appear as the oscillations
 of $r$ and $q$ because
 $\delta\phi = 2q.{\delta}r + 2{\delta}q.r$.    
 We have phonons when $q = q_o$, and omons ( a new kind of quantum 
quasiparticle representing a
phononlike wave of the {\bf o}scillations of {\bf m}omentum )  when 
$r = d$.  We note that a system like liquid ${^4He}$ is expected to exhibit:
 (i) no transeverse mode because the
shear forces between particles are negligibly small, and (ii) only one
branch of longitudinal mode because the system is isotropic.  Evidently, 
${\omega}_r({\rm Q})$ of phonons 
can be represented, to a good approximation, by the dispersion of 
the elastic waves in a chain of
identical atoms and it can be obtained from Eqn.(13)  
by replacing $\beta$ by $m$ and $C$
 by C  $[= 4{\pi}^2.C/{d}^2 =
 2{\pi}^2.k_BT_o/{d}^2]$. 
 However, for a better accuracy $d$ and C 
 should, respectively, be considered descending and ascending
 functions of Q because  
 increase in the energy of particles affected by 
 an excitation reduces  
  WP size and this renders a decrease in $d$ and
 an increase in C.  As such   
 ${\omega}_r({\rm Q})$ should be expressed as   
$${\omega}_r({\rm Q}) = 
 {\sqrt{4{\rm C(Q)}/m}}.|\sin({\rm Q}d{\rm (Q)}/2)|
 \eqno (14).$$
\noindent
which also reveals that the phase and group velocities of low Q
 modes are  
$$v_p = v_g = {\sqrt\pi}h/2md   \eqno (15)$$ 
\smallskip
The energy dispersion E(Q) should be    
phononlike till the excitation wavelength 
${\Lambda} > d$. However, since  
 the momentum and energy of an excitation of ${\Lambda} < \sigma$
 is shared only by a single atom, we expect  
E(Q) = ${\hbar^2{\rm Q}^2/2{m}^*}$ (with $m^*$ = 
effective mass of an atom).  Obviously, the transition of E(Q) from
 phononlike curve to E(Q)=${\hbar^2{\rm Q}^2/2{m}^*}$ takes place over
 the range, $2\pi/d < {\rm Q} < 2\pi/{\sigma}$.  Evidently, 
 E(Q) does not touch  the Q-axis at any Q $\ne 0$.  Its  
 maxon point lies at $\approx \pi/{\sigma}$
 and the roton minimum around the middle point
 of $2\pi/d$ and $2\pi/{\sigma}$.  As a whole  
E(Q) represents Landau type spectrum.  The E(Q) of N-phase is, 
obviously, expected to differ from this spectrum.   
\smallskip 
To obtain ${\omega}_q({\rm Q})$ of omons, we note 
 that the equation of motion of $r_s$ (the $r$ of
$s$-th atom),
{\it i.e.},  ${{\partial}^2_t}r_s  =
 - {\frac{1}{4}}.{\omega}_o^2.[2r_s - r_{s-1} - r_{s+1}]$,
  transforms into a  similar equation for $p_s = \hbar{q_s}$ by  
operating $m.{\partial}_t$.  This reveals 
${\omega}_r({\rm Q}) = {\omega}_q({\rm Q})$.  As concluded in 
Section 5.4, an omon 
is an anti-phonon quantum quasiparticle.     
\smallskip
The anomalous behaviour of ${\omega}_r({\rm Q})$ 
at low Q arises due to Q dependence of
 C and $d.$  Using this observation to explain 
the anomalous nature of E(Q) of $He-II$ [14],     
  we found that moderately Q dependent  
 $d$ and C become nearly Q independent when Q approaches $\pi/{\sigma}$. 
  It seems obvious because $d \not < \sigma$.             
\smallskip
Considering $\psi = \sum_if(r_i)\phi$ as the wave function of
 an excited state and $\phi$ that of ground state, Feynman [15] showed 
that the excited state energy is minimum for $f(r_i)
 = \exp(i{\bf k}.{\bf r}_i)$ and E(Q) =
${{\hbar}^2{\rm Q}^2}/2m{\rm S(Q)}$ (with S(Q) = structure factor); this  
 E(Q) for $He-II$ is, however,  
  found to be about two times of the 
 experimental value.  Feynman and Cohen [16] later obtained better 
 results but with considerable discripancy at higher Q.  Since
 our state function 
({\it cf.} Eqn.(10) and (6))
 satisfies all considerations of Feynman, 
we obtained E(Q) =${{\hbar}^2{\rm Q}^2}/4m{\rm S(Q)}$ by using the 
fact that a SMW pair (rather than a single particle) forms   
 the basic unit of the system until Q $> 2\pi/{\sigma}$.  This 
explains the observed E(Q) for $He-II$
 to a good accuracy [14].  
\bigskip
\centerline{\bf 5.3 Two Fluid Behaviour} 
\bigskip
Since ${\phi}_N^o(q_o)$ and ${\phi}_N^e(K)$ deal separately 
with $q$ and $K$ motions of particles they represent 
 two different components in the system.  The component
${\phi}_N^o(q_o)$ representing particles in their ground state, obviously,
has zero entropy (S=0); this also has $\eta = 0$ because  
particles are constrained to move in the order of
their locations (${\it cf.}$ Section 5.1).
  In the rotating fluid, however, particles moving on the neighbouring
cocentric  circular paths have relative velocity and the system shows its
natural viscous behaviour. The excitations are  
the effects that can propagate from one
end of the system to the other against the cosely packed 
WPs in the background.  They, obviously, form a kind of gas
(as envisaged by Landau [17]) that accounts for the total S and other
thermal properties of the system.    
  They also render a $\eta \ne 0$ because their effects can lead to    
frictional movement of particles.  As such ${\phi}_N^o(q_o)$ has
 the basic properties of
S-fluid, while ${\phi}_N^e(K)$ has those of N-fluid; however, 
 these fluids are inseparable since each 
particle participates in ${\phi}_N^o(q_o)$
 as well as ${\phi}_N^e(K)$.       
   All these aspects vindicate two fluid theory of Landau [17].       
\bigskip
\centerline{\bf 5.4 Energy Gap and Self Energy} 
\bigskip
With $T$ moving below $T_{\lambda}$, the WPs of neighbouring
atoms have an overlap because $\lambda/2$ tends to increase by 
pushing particles to have increased $d$ against 
interparticle attraction.  This perturbs ${\psi}^{\pm}$
and its energy $\varepsilon_o$.  Considering a pair of particles and
following the standard method dealing with similar situation (e.g. the
formation of a molecular orbital from two identical atomic orbitals), we
find that the energy of perturbed state becomes
  $\varepsilon_o \pm |v|$; here $|v|$ is the
 expectation value of interatomic attraction.  $|v|$ could better be 
replaced by $|v(T)|$ as the overlap may depend on $T$.  The state of lower
energy $(\varepsilon_o - |v(T)|)$ represents a kind of {\it bonding (or
paired) state} while that of $\varepsilon_o + |v(T)|$ an {\it antibonding (or
unpaired) state}.  The lower energy state can thus  be regarded as a state of 
bound pairs.  Since this happens to all
particles, their ground state energy falls
 from $N\varepsilon_o(T_{\lambda})$ at $T =
T_{\lambda}$ to a new value $N\varepsilon_o(T)$ = 
 $N\varepsilon_o(T_{\lambda}) - N|v_N(T)|$ =
$N\varepsilon_o(T_{\lambda}) - E_g(T)$ at $T<T_{\lambda}$; note 
that $|v_N(T)|$ may differ from $|v(T)|$
obtained for an isolated pair. This gives 
$$E_g(T) = N[\varepsilon_o(T_{\lambda}) - \varepsilon_o(T)] \approx 
Nh^2(d_T - d_{\lambda})/4md^3_{\lambda} 
\eqno(16) $$
\noindent
$E_g(T)$ can be identified as : (i)
an order parameter of the transition, (ii) a {\it collective
binding among all particles} (clearly different from the binding 
of two electrons in a Cooper pair) rendering the system to behave like a 
single molecule [18], and (iii) an energy gap between 
S and N phases in a sense that S phase becomes N phase if
$E_g(T)$ energy is supplied from outside.  The system retains fluidity
since $|v_N(T)| << \varepsilon_o$.  
\smallskip
To determine the factor that controls the overlap of neighbouring 
WPs and hence the value of $E_g(T)$, we note that the increased 
 size of WPs in the bonding state forces the system to expand. 
  This requires energy (to be managed from
 within the system ) to work against inter-atomic
 attraction.  In this context we find that the system at
 $T_{\lambda}$ has thermally excited $N^*(T_{\lambda})$ particles  
devoid of quantum correlation for their $\Lambda$ being 
  $< \sigma$.  The transition of such particles
 to their ground state 
  releases an additional $\Delta{\epsilon}(T) = [N^*(T_{\lambda}) 
 -N^*(T)].k_BT_o{\ln{2}}$ energy when 
 the system is cooled from $T_{\lambda}$ to $T$.  This is because
 the lower energy level of the transition lies 
 below the zero line of the kinetic energy by -$k_BT_o.{\ln{2}}$.  
  Evidently, the desired controlling factor is
 $\Delta{\epsilon}(T)$ energy available for the 
expansion. 
\smallskip
 The process of expansion 
 pushes the particles 
 to their higher potential energy rendering a  
net increase by $\Delta{V_s(T)}$ 
 termed as self energy.  Naturally, $\Delta{V_s(T)}$
 = ${\Delta\epsilon{(T)}}$.  
In this state of new equilibrium the zero-point 
repusion $(=h^2/4md^3)$ equalises with interparticle attraction. 
This gives $\Delta{V_s(T)}$ = $|E_g(T)|$.    
 We note that: (i) $\Delta{V_s(T)}$ is a function of $q$ because 
it depends on the size of WPs, (ii) it provides a basis for    
 the propagation of omons in the S-phase, and (iii) it can be  
 recognised as the energy of omon field.  Since
  $\Delta{V_s(T)}$ increases with decreasing $T$, 
 it implies that omon field intensity increases  
when phonon field intensity decreases and vice versa. Evidently, 
$\Delta{V_s(T)}$ could either be identified as the energy of phonon field 
absorbed by the particles or the   
 omon is considered as an anti-phonon quantum quasi-particle. 
Since $\Delta{V_s(T)}$ attains its maximum value at $T=0$, it   
can serve as a source of energy  
for creating collective motions even at $T=0$. 
\bigskip
\bigskip
\centerline{\bf 5.5 Consequences of Energy Gap} 
\bigskip
If two heads $X$ and $Y$ in the system 
have small $T$ and $P$ (pressure)
 differences, the equation of state is     
$E_g(X) = E_g(Y) + {\rm S}{\Delta}T - {\rm V}{\Delta}P$.   
Using $E_g(X) = E_g(Y)$ for equilibrium, we get  
$${\rm S}{\Delta}T = {\rm V}{\Delta}P     \eqno(17)$$  
\noindent
This reveals that the system should exhibit thermo-mechanical and
mechano-caloric effects.  This also concludes that measurement 
of $\eta$ by capillary flow method perfomed under 
the condition  ${\Delta}T = 0$ and that of thermal conductivity $(\Theta)$
performed under ${\Delta}P = 0$ should reveal  
$\eta = 0$ and $\Theta \approx \infty$.  This explains why    
$He-II$ is a superfluid
of infinitely high $\Theta$. 
\smallskip
Equating $E_g(T)$ with the energy of flowing system 
we obtain the upper bound
 of {\it critical velocity} for which
the S-phase becomes N-phase.  We have 
$$v_c(T) = \sqrt{[2E_g(T)/Nm]} \eqno (18) $$
\noindent 
The lower $v_c$, at which the superfluid may show signs of viscous
behaviour, arises due to creation
of quantized vortices, but this does not destroy 
superfluidity.  Similarly, we find the following relation for 
{\it coherence length} (not to be confused with healing length [4]), 
$$\xi = h.{\sqrt{[N/2mE_g(T)]}} \eqno (19) $$ 
\smallskip
Correlating the {\it superfluid density}, ${\rho}_s$, as the 
order parameter of the transition, with $E_g(T)$ we find a 
new relation  
$${\rho}_s(T) = [E_g(T)/E_g(0)]{\rho}   \eqno (20) $$
\noindent
to determine  
${\rho}_s$ and {\it normal density},
 ${\rho}_n = {\rho} - {\rho}_s$.    
 \bigskip
 \centerline{\bf 5.6 Quantized Vortices}
 \bigskip
Using the symmetry property of a state of
 bosons, Feynman [15] showed that
the circulation, $\kappa$, of the velocity field should 
be quantized, ${\it i.e.}$, 
$\kappa = \frac{nh}{m}$ with $n = 1,2,3,...$
However, Wilks [1] has rightly pointed out that this account does not
explain the fact that $He-I$ to which Feynman's
argument applies equally well, does not exhibit quantized vortices.  We 
find that particles in S-phase maintain phase coherence, 
not existing in N-phase, and for this reason the quantized 
vortices can be observed in S-phase only.    
 \bigskip
 \centerline{\bf 5.7 Single Particle Density Matrix and ODLRO}
 \bigskip
Using Eqns.(10), and (6), we have [8] the single particle density matrix, 
$$\rho({\bf R^*-R}) = \left[ \frac{N(o)}{\rm V} + 
\frac{N}{{\lambda}'^3_T} .\exp{\left(- 2\pi.
\frac{|\bf {R^*-R}|^2}{{{\lambda}'_T }^2} \right)} \right].
\left({\frac{2}{\rm V}}.
\cos^2[\frac{\pi(r''-r')}{d}]\right)  \quad \quad
   \eqno(21)$$
\noindent
where ${\lambda}'_T = h/{\sqrt{2\pi(4m)k_BT}}$ 
represents thermal wavelength attributed to $K$ motions.  
We used ${\bf q}_o.({\bf r''} -{\bf r'})^* =  2n{\pi} + 
{\bf q}_o.({\bf r''} -{\bf r'})$.  The term in big (..) represents 
the variation of density over a single AR.  
The  fact that under the limit $|{\bf R^*} - {\bf R}|$
 tends to $\infty$, $\rho({\bf R^*-R})$ has 
nonzero value for $T<T_{\lambda}$ and zero for  
$T \ge T_{\lambda}$ since $N(o)$ (= number of particles
representing  the SMW pairs of $K=0$) is $\ne 0$ for $T<T_{\lambda}$ and
$=0$ for $T \ge T_{\lambda}$.  It clearly proves the existence of off
diagonal long range order (ODLRO) in the S-phase of the system.
\bigskip
\centerline{\bf 5.8 Logarithmic Singularity of Specific Heat}
\bigskip
The specific heat $C_p(T)$ of the system is expected to show usual cusp
at $T_{\lambda}$ if BEC of SMW pairs is considered as the only 
mechanism of the transition.  But  
our system at its $\lambda$-point also has an
 onset of ordering of particles in phase 
 space rendering widely different 
changes in ${\phi}'$-position of different particles.  To determine the 
 corresponding change in energy $\Delta{\varepsilon}$ we, however, 
 assume for simplicity that of the $N^*(T_{\lambda})$ uncorrelted particles
in their excited states, $N_{\lambda}$
 make significant contribution to $\Delta{\varepsilon}$ 
 and they move from their ${\phi}'= 
 2n{\pi} \pm (\pi - \delta{\phi}_{\lambda})$ at $T^+_{\lambda}$ 
(just above $T_{\lambda}$) to $ {\phi}' = 2n{\pi}$ at
$T^-_{\lambda}$ (just below $T_{\lambda}$).  This gives     
$$\Delta{\varepsilon} = - N_{\lambda}k_BT_o\left[{\ln{2.\cos^2{\left(\frac
{2n{\pi} \pm (\pi - \delta{\phi}_{\lambda})}{2}\right)}}  - \ln{2}}\right]
 \eqno(22) $$    
\noindent
Following the
 theories [19] of critical phenomenon we may define 
$$\delta{\phi}_{\lambda} = \delta{\phi}_{\lambda}(o).|\zeta|^{\nu}
\left[ 1 + a_2.|\zeta|^2 + a_3.|\zeta|^3 \right] \eqno(23)$$     
\noindent
with $\zeta = \frac{T-T_{\lambda}}{T_{\lambda}}$.  
To a good approximation we have  
$$\Delta{\varepsilon} = -
 N\left(\frac{T-T_{\lambda}}{T_{\lambda}}\right).
k_BT_o.\ln{\left( \frac{\delta{\phi}_{\lambda}(o).
|\zeta|^{\nu}}{2}\right)^2}
  \eqno(24)$$ 
\noindent
by using $\delta{\phi}_{\lambda} =
 \delta{\phi}_{\lambda}(o).|\zeta|^{\nu}$ and 
$N_{\lambda} = N.\frac{T-T_{\lambda}}{T_{\lambda}}$; the latter  
 expression is so chosen to ensure that $\Delta{\varepsilon}$ 
 does not diverge at $T_{\lambda}$ and it decreases with decreasing $T$
 through $T_{\lambda}$.  Eqn.(24) gives
$$C_p(T \approx T_{\lambda})
 \approx - \frac{N}{T_{\lambda}}.k_BT_o.[2\nu.\ln|\zeta| 
+ \ln{(\delta{\phi}_{\lambda}(o)^2)}
- \ln{4} + 2{\nu}] \eqno (25)$$
\bigskip
\centerline{\bf 5.9 Properties of $He-II$}
\bigskip
All important conclusions of our theory that the S-state of the system
should exhibit: (i) two fluid behaviour, (ii) negative volume
 expansion, (iii) ${\eta} = 0$ for
 its capillary flow, (iv)
 ${\eta} \ne 0$ in the state of its rotation,
 (v) infinitely high $\Theta$,
(iv) quantized vortices, (v) coherence of particle motion,
{\it etc.}, are well known properties of
 $He-II$.  That our theory  
provides an accurate account of
 the thermodynamic and
hydrodynamical properties of $He-II$ is well evident since  
its theoretical E(Q), ${\rho}_n(T)$ and ${\rho}_s(T)$ 
 [9,14] match closely with experiments.
  Our estimates [20] of :
 (i) $T_{\lambda} 
\approx 2.03{\rm K}$, (ii) upper bound $v_c$ varying
 smoothly from 0 (at
$T_{\lambda}$) to $\approx 9 {\rm m/sec.}$ (at $T=0$),
  (iii) anomalous variation of $v_p$ and $v_g$ and their low Q value 
$\approx 227 {\rm m/sec.}$, (iv) $\xi$ varying
 smoothly from $\approx 10^{-6} {\rm cm}$,
 at $T=0$ to infinitely large value at $T=T_{\lambda}$,
${\it etc.}$ agree with experiments (see [4] for (i-ii), [6] 
for (iii), and [21] for (iv)).  
\smallskip
For the first time our theory explains the origin of the  
experimentally observed logarithmic
 singularity in $C_p(T)$ at $T_{\lambda}$.  
Using the parameters of liquid ${^4He}$ in Eqn.(25), we have
$$C_p({\rm J/mole.K}) \approx
 -5.24.\ln|\zeta| - 9.5 = A.\ln|\zeta| + B  \eqno(26)    $$
\noindent
where we use $\nu = 0.55$ and $\delta{\phi}_{\lambda}(o) = \pi$, while 
the experimental results reveal
A = 5.355 and B = -7.77 for $T > T_{\lambda}$ and A = 5.1 and B = 15.52
for $T < T_{\lambda}$ [22].  The fact that our A value agrees closely with 
experiments speaks of the accuracy of our theoretical result.  With 
respect to our choice of $\nu = 0.55$ and
 $\delta{\phi}_{\lambda}(o) = \pi$, we note that:
 (i) $\delta{\phi}_{\lambda}$ originates basically from change
 in momentum ${\Delta}k \approx {\xi}^{-1}$ and $\xi$ varies 
around $T_{\lambda}$ as $|T - T_{\lambda}|^{-\nu}$; note that 
critical exponent $\nu$ lies in the range 0.55 to 0.7.  While we have 
no definite reason for our choice of
 $\delta{\phi}_{\lambda}(o) = \pi$, however, this is the largest possible 
value by which phase position of a particle can change.   
\smallskip
While our value ($ \approx $ - 21 K) of potential energy
 per ${He}$ atom does not differ from others [12], our zero
 point kinetic energy (${\approx}$ 3.93 K) is much 
lower than ${\approx}$ 14 K estimated by others [12].  Evidently,
the configuration revealed by our theory is energetically
favourable one.  Using the values of $\xi$ and $E_g(T)$
 and following the standard method [23] we obtained  
the time of persistent currents for $He-II$ as 
$\approx 10^{34}$sec. which is 
much larger than the life of our universe,
 {\it i.e.} $\approx 10^{18}$sec.  
\bigskip
\centerline{\bf 5.10 Low Dimensional System}
\bigskip
A 1-D (or 2-D) system that could be realised in a laboratory 
 is nothing but a 3-D 
system having 2 (or 1) dimension(s) reduced to the size of the 
order of $d$ (or $\lambda/2$); the $k, K, r$ and $R$, obviously, 
 retain their 3-D character.
  The only important aspect in which these systems differ from 
a real 3-D system is that $k$ can be controlled by the smallest 
dimension of these systems.  Our theory, obviously, has no difficulty in
 explaining the superfluid behaviour of such systems
because contrary to the convetional theories presuming particles to have 
$\lambda \approx \infty$, it presumes $\lambda/2 \le d$ and  
 finds $\lambda/2 =d$ as the basic condition of superfluidity. While 
particles in a low dimensional system
 can not have $\lambda \approx \infty$ (characteristic of $p=0$ 
condensate), they 
can certainly satisfy $\lambda/2 =d$ if the density of
 particles is high enough to satisfy $d \le t$
 where $t$ represents the diameter of 
 the narrow channel (or the thickness of the film); here $d$ 
and $t$ are defined by excluding the 
solidified atomic layers on the surface of the channel or the substrate.
  If $t < d$, We can only have $\lambda/2 \le t$ since $\lambda/2$ is now 
 controlled by $t$.   
Naturally, $\lambda/2$ is always $< d$ and  
 the system would not exhibit superfluidity.  For this case 
 we also find that the    
WPs of neighbouring particles fail to connect   
each other to develop a macroscopic chain (or 2-D network) of SMWs. 
 Evidently, $d \le t$ identified as the
 {\it connectivity condition} 
 has to be satisfied by the system for superfluidity.  In other words 
 the particle number density for a given $t$ of 1-D (or 2-D) system
should be $> t^{-1}$ (or $t^{-2}$).  
\smallskip
 Note that: (i) the surface (in contact with the sample)
 of channels (or substrates) used to 
realise 1-D (or 2-D) systems are very rough at the scale 
of $d$, (ii) the diameter of the channels has fluctuations 
of the order of $d$, and (iii) particles moving near the
 substrate surface have increased effective mass because they experience 
surface potential of attractive nature.  Naturally, the $\lambda$-
 transition in different parts of the sample occurs at
 different $T_{\lambda}(d) < T_{\lambda}$ of the bulk system.  
Further since the system attains maximum density at $T > T_{\lambda}$, 
and superfluidity can be observed only when the process of 
transition is completed in the entire sample, it is evident that 
$T$ of : (i) the peaks of maximum density, ($T_m$)
and specific heat ($T_{sh}$), and (ii) the
 onset of superfluidity, ($T_{sf}$) would satisfy 
$T_m > T_h > T_{sf}$ which agrees with experiments [24]. The locations 
of these temperatures would also not be so well defined as 
$T_{\lambda}$ of the bulk ${^4He}$.  Note that this analysis 
does not apply to a real case of 1-D (or 2-D) systems in which 2 (or 1) 
dimension(s) are reduced to zero exactly.           
\bigskip    
\centerline{\bf 5.11 Similarity with Photons in Laser Beam}
\bigskip
We note that the system below $T_{\lambda}$ defines a 3-D network 
of SMWs extending from one end to the other without
 any discontinuity.  In lasers too these are the standing waves 
of electromagnetic field that modulate the probability of 
of finding a photon at a chosen phase point.  The basic difference 
between the two lies in the number of bosons in a single AR.  In case of 
lasers this could be any number since photons are non-interacting but 
for the systems like ${^4He}$ and ${^{87}Rb}$ one AR can only one atom.
\bigskip
\centerline{\bf 6. CONCLUDING REMARK}
\bigskip
Our theory is consistent with the volume exclusion
 condition [25], as well as 
(ii) the microscopic and macroscopic
uncertainty [4].  It vindicates:
 (i) the two fluid theory of Landau [17], (ii) London's idea of
macroscopic wave function of the S-state [26], (iii) the
observation of Bogoliubov [27] that superfluidity is an interplay of
interparticle interactions, {\it etc}.  The theory, however,  
 does not reveal the existence of $p=0$ condensate.  When the  
interactions are switched off, a particle does not experience
 the existence of the other. 
The states of the system are, obviously, 
 single particle states and   
its ground state is defined by $K=0$ and $q_o = \pi/L (\approx 0)$;
 $L=$ size of the container. 
  Using $q = 0$ alongwith $\alpha = 0$, we note that Eqn.(6)
becomes the usual wave function of plane waves.  It is evident 
that our theory has the basic foundation that    
should help in explaining 
the behaviour of 1-D and 2-D analogs of our system 
and that of ${^{87}Rb}$ [28] like samples.  Our preliminary 
results [14] corroborate this point.  
\smallskip
In order to extend our formulation to a fermi system, we can use 
 Eqn.(6) by noting that :(i) in view of the condition $\lambda/2 \le d$
  all HC fermions     
in their ground state can have $q= \pi/d$ and (ii) they need to follow 
fermi distribution over the possible states of  
 $K$ representing a free particle of mass $2m$
 placed in a 3-D box; note that a SMW pair is the basic unit of the system 
unless the excitation wavelength $\Lambda < \sigma$ in which 
 case the excitations should be single particle excitations.  We may
 require to replace 
$m$ by effective mass $m^*$ due to interparticle interaction. In
  case of an atomic nucleus
 the allowed values of $K$ would 
 correspond to a free particle of mass $2m^*$ 
kept in a spherical shell; the shell may have some deformations 
in its shape depending on its volume or related parameters.  
   As such our theory finds no 
difficulty in identifying an atomic nucleus as a tiny drop of fermi 
superfluid.  In view of our inferences ({\it c.f.} Section 5.1) 
 which are applicable HC bosons and fermions alike 
 this drop should represent  a close packing of the WPs of 
 different nucleons, obviously, forbidden to have  
 random motion; they should also have an orderly arrangement in 
phase space.  These new inferences should certainly help in developing a 
better understanding of the physics of an atomic nucleus.  For example 
 the origin of collective oscillations of its nucleons as well as the 
 single particle excitations can be better conceived.  Using 
these observations we are 
working out the basic aspects of these systems and we hope
 to report our findings soon.         
\smallskip
As such we hope that our theory would help
 in developing a unified model  
of superfluidity and superconductivity and a better 
understandinng of the physics of atomic nucleus and neutron star.
\vspace{20cm}
\centerline{\bf REFERENCES}
\vspace{.3cm}
{\leftskip=.7cm
{\parindent=-.5cm
{\bf 1.} J. Wilks, {\it The properties
 of Liquid and Solid Helium},
(Clarendon Press, Oxford, 1967).
\par}
{\parindent=-.5cm
{\bf 2.} Z.M. Galasiewics, {\it Helium 4}, (Pergamon, Oxford, 1971)
\par}
{\parindent=-.5cm
{\bf 3.} A.D.B. Woods and R.A. Cowley, Rep. Prog. Phys. {\bf 36},
1135(1973)
\par}
{\parindent=-.5cm
{\bf 4.} S.J. Putterman, {\it Superfluid Hydrodynamics}, (North-Holland,
Amsterdam, 1974)
\par}
{\parindent=-.5cm
{\bf 5.} K.H. Benneman and J.B. Ketterson, Eds., {\it The Physics of
Liquid and Solid Helium}, Part-I, (Wiley, New York, 1976).
\par}
{\parindent=-.5cm{\bf 6.} H.R. Glyde and E.C. Svensson, in {\it Methods
of Experimental Physics}, edited by D.L. Rice and K. Skold, Neutron
 Scattering, {\bf Vol. 23}, Part B,
 (Academic, San Diego, 1987), pp 303-403.
\par}
{\parindent=-.5cm{\bf 7.}
 A. Griffin, D.W. Snoke and A. Stringari, Eds., {\it {Bose Einstein
Condensation,}}  (Cambridge University Press, Cambridge, 1995).
\par}
{\parindent=-.5cm{\bf 8.}
 J.G.M. Armitage and I.E. Ferquhar, Eds., {\it {The Helium Liquids,}}
  (Academic Press, London, 1975).
\par}
{\parindent=-.5cm{\bf 9.} Y.S. Jain, {\it Microscopic Theory of
 Superfluidity of a System Interacting 
Bosons : The Liquid $^4He$}, Techn. Rep. No. SSP/2 (1995).
\par}
{\parindent=-.7cm{\bf 10.}
 Y.S. Jain in (a) {\it Proc. of XX International Workshop on 
Condensed Matter Theories}, Pune University Pune,
 Dec. 9-13, 1996 (to be published) and 
(b) {\it Proc. of International Workshop on High Temperature 
Superconductivity}, Rajasthan University Jaipur, Dec. 16-21, 1996 
(to be published)
\par}
{\parindent=-.7cm{\bf 11.} K. Huang and C.N. Yang, Pys. Rev. {\bf 105}, 
767(1957).
\par}
{\parindent=-.7cm{\bf 12.}
C.-W. Woo in Ref.[5], pp 349-501.
\par}
{\parindent=-.7cm{\bf 13.}
G. Taubes, Sience {\bf 269}, 152 (1995).
\par}
{\parindent=-.7cm{\bf 14.} Y. S. Jain and Coworkers (unpublished work),
 and D. R. Chowdhury, Ph.D. Thesis, Department of Physics,
 North-Eastern Hill University (1996).
\par}
{\parindent=-.7cm{\bf 15.}
R.P. Feynman, in {\it Progress in LOw Temperature Physics},
edited by C.J. Groter, 
 (North-Holland, Amsterdam, 1955), {\bf Vol. 1}, p. 17
\par}
{\parindent=-.7cm{\bf 16.}
R.P. Feynman and M. Cohen, Phys. Rev. {\bf 102}, 1189(1956).
\par}
{\parindent=-.7cm{\bf 17.} L. D. Landau, J. Phys.
 (USSR) {\bf 5}, 71 (1941) and {\it ibid} {\bf 11},
 91 (1947); Reprinted in english in Ref. [2] pp 191-233, pp 243-246.
\par}
{\parindent=-.7cm{\bf 18.}  
C.J. Foot and A.M. Steane, Nature {\bf 376}, 213 (1995).
\par}
{\parindent=-.7cm{\bf 19.}  
M.E. Fisher, Rep. Prog. Physics {\bf 30}, 615 (1967).
\par}
{\parindent=-.7cm{\bf 20.}
To a good approximation, the system being a close packed
 arrangement of WPs could be assigned a symmetry such
 as hcp, fcc, {\it etc.}  Naturally, $d$ can be as large as
$1.122a$ for fcc arrangement and as small as $a = {(V/N)}^{1/3}$
for sc.  In our estimates we use $d = 1.091a$ of bcc arrangement. 
\par}
{\parindent=-.7cm{\bf 21.}
The $\xi$ (Eqn. 19) should be related to the spatial separation 
between two particles in SMW configuration 
keeping definite phase correlation and it appears to 
be the origin of: (i) nearly constant thickness
  ($\approx 3{\rm x}10^{-6}$  
cm) of the film observed in 
the beaker film flow experiment
 (see L.J. Campbell in Ref. [8], Chapter 4), 
(ii) the radius of vortex rings ($\approx 5{\rm x}10^{-6}$
 to $10^{-4}$ cm)
observed for vortices created by moving ions 
(see Ref. [1], Chapter 12), {\it etc.}
\par}
{\parindent=-.7cm{\bf 22.}
G. Ahlers in Ref. [5], Chapter 2. 
\par}
{\parindent=-.7cm{\bf 23.}
C. Kittel, {\it Introduction to Solid State Physics}, (Wiley Eastern,
New Delhi, 1976)
\par}
{\parindent=-.7cm{\bf 24.}
A. Eggington in Ref.[8] p. 195.
\par}
{\parindent=-.7cm{\bf 25.} 
P. Kleban, Phys. Lett. {\bf 49A}, 19(1974).
\par}
{\parindent=-.7cm{\bf 26.} 
F. London, Phys. Rev. {\bf 54}, 947 (1938). 
\par}
{\parindent=-.7cm{\bf 27.}
N. N. Bogoliubov, J, Phys. (USSR) {\bf 11}, 23 (1947); Reprinted
 in english in Ref.[2], pp 247-267.  
\par}
{\parindent=-.7cm{\bf 28.}
M.H. Anderson, J.R. Ensher, M.R. Mathews, C.E. Wiechman 
and E.A. Cornell, Science, {\bf 269}, 198(1995)
\par}}
\end{document}